\documentstyle[aps,preprint,psfig,epsf]{revtex}

\newcommand{\be}{\begin{eqnarray}}
\newcommand{\ee}{\end{eqnarray}}

\newcommand{\spett}{P_{A}^N(k,E)}
\newcommand{\qud}{Q^{2}}
\newcommand{\hd}{^{2}H}

\newcommand{\heq}{^{4}He}
\newcommand{\cd}{^{12}C}
\newcommand{\fec}{^{56}Fe}
\newcommand{\sez}{\frac{d^{2}\sigma_{eA}}{d{\Omega}d{\nu}}}
 \newcommand{\sezqe}{\frac{d^{2}\sigma_{eA}^{qe}}{d{\Omega}d{\nu}}}

\begin{document}

\title{INCLUSIVE ELECTRON SCATTERING FROM NUCLEI: $y$ SCALING AND FINAL
STATE INTERACTION}

\author{ Dino FARALLI}

\address{Department of Physics, University of Perugia, and \\ Istituto
Nazionale di Fisica Nucleare, Sezione di Perugia \\
 Via A. Pascoli,
I--06100 Perugia, Italy}

\author{ Claudio CIOFI degli ATTI
\thanks{Permanent address:
Department of Physics, University of Perugia, and Istituto
Nazionale di Fisica Nucleare, Sezione di Perugia,
 Via A. Pascoli,
I--06100 Perugia, Italy
.} and
 Geoffrey B. WEST}
\address{Theoretical Division, T-8, MS B285, Los Alamos National
Laboratory\\ Los Alamos, NM 87545, USA .}

\date{\today}


\maketitle
\begin{abstract}
The recent TJLAB experimental data on
inclusive electron scattering at high momentum transfer from complex nuclei
are analyzed in terms of $y$ scaling, taking  into account the
final state interaction (FSI) of the struck, off-shell nucleon. It is shown
 that
at large negative values of $y$ ($x>1$), the $Q^2$ dependence of the
FSI  is mostly  driven by the {\it elastic} nucleon-nucleon cross section,
and that,
as a result, the scaling function decreases with $Q^2$, in agreement
with  experimental data.
 \end{abstract}

\section{Introduction}

The recently released TJNAF E89-008 experimental data on inclusive
electron scattering $A(e,e^{\prime})X$  at $x>1$ ($x=\frac {Q^2}{2M\nu}$ is
the Bjorken scaling variable) and high momentum transfer \cite{arri}
could provide relevant information on  nucleon momentum
distributions in nuclei  and the mechanism of final state interactions  in
inclusive processes. In this contribution, these data are analyzed
by calculating both the cross sections and the $y$-scaling functions,
using, in the latter case,  a recently proposed new approach \cite{CW}.
 The effects of FSI will  be considered, stressing
  the relevant role
 played by the off-shell kinematics for the struck nucleon
  before its rescattering from spectator nucleons in the medium.

\section{The Inclusive Cross Section}

The inclusive electron-nucleus cross section can be written in the following
 general form:
\be
\sez=\sigma_{Mott}\left[ W_{2}^{A}(Q^{2},\nu )+2\tan^{2}(\theta /2)W_{1}^{A}(Q^{2},\nu
)\right]
 \label{sez}
 \ee
 where $Q^{2}={\bf q}^{2}-\nu^{2}$  is the four-momentum transfer squared,
 and $\sigma_{Mott}$  the Mott cross section.
In plane  wave impulse approximation (PWIA),  the nuclear structure functions
$W_{1,2}$ are given by

\be
W_{i}^{A}(Q^{2},\nu) =\sum_{N=1}^{A} \int d{\bf k}\int dE \,P^{N}({k},E)
\left[
C_{i}W_{1}^{N}({Q^2},\nu\prime)+D_{i}W_{2}^{N}({Q^2},\nu\prime) \right]
\label{struct}
\ee
where $i=\{1,2\}$,
$W_{1(2)}^{N}$ are   the nucleon structure functions, $C_{i}$ and
$D_{i}$ are some kinematics factors, $\nu^{\prime}=(p\cdot Q)/M$ where $p$ is the
four-momentum of the struck off-shell nucleon, and  $P^{N}({k},E)$ its
{\em spectral function}; we denote the  three-momentum
of the struck nucleon by $k=|\bf p|$  and  its removal energy
by
$E = E_{min} + E^{\ast}_{A-1}$  where $E_{min}=M_{A-1}+M-M_{A}$.

 We have calculated the cross section (1) both in the quasi-elastic
 and inelastic regions using the experimental
nucleon structure functions and the spectral
 function  from Ref. \cite{CS}. The PWIA results for
 $^{56}Fe$ are represented in
 Fig. 1 by the dotted curve, which is the sum of the quasi-elastic
  (short dashes) and inelastic (long dashes) contributions. The well-known result, that the PWIA
underestimates the cross section at low values of $\nu$ ($x>1$) is shown to hold true for the new
data. We have  included  the FSI of the struck nucleon using the method of Ref.
\cite{CS}. The results are shown by the full curve where it is seen that the agreement
with the experimental data is satisfactory. More details on the
calculation of the FSI will be given in  Section IV.  At $x>1$,
the contribution from inelastic channels is always very small, the
dominant process being  quasi-elastic scattering; this therefore justifies
 an analysis of
the data in terms of $y$-scaling.

\section{Quasi-elastic scattering and {$y$}-scaling}

In PWIA the quasi-elastic cross-section is given by

\be
\sezqe
 =2\pi \sum_{N=1}^{A} \int\displaylimits_{E_{min}}^{E_{max}(q,\nu)}
 d\!E\int\displaylimits_
 {k_{min}(q,\nu,E)}^{k_{max}
(q,\nu,E)} k\,d\!k\, \spett \sigma_{eN}(Q^2,\nu^{\prime}) \frac{E_{p}}{q}
\label{sezpw2}
 \ee

\noindent where $E_{p}=\sqrt{M^{2}+({\bf k}+{\bf q})^{2}}$ is the {\it on shell}
 energy of the struck nucleon after photon absorption , $q=|{\bf q}|$ and $\sigma_{eN}$ is the
elastic  electron-nucleon cross section.
 The integration limits in (\ref{sezpw2}) are determined from
energy conservation:

\be \nu+M_{A}=\sqrt{M^{2}+({\bf k}+{\bf q})^{2}}+\sqrt{M_{A-1}^{\ast 2} +{\bf k^{2}}}.
\label{consen}
\ee
\noindent where $M_{A-1}^{\ast}=M_{A-1}+E_{A-1}^{\ast}$ is the mass of the excited $(A-1)$
system.
At large values of the momentum transfer, the following relation holds

\be
 \sezqe = F_{A}(q,\nu) \left[(Z\sigma_{ep}+N\sigma_{en})
\frac{E_{p}}{q}\right]_{(k=k_{min},E=E_{min})}
  \label{sezpw3}
\ee
\noindent where the nuclear structure function,  $F_{A}(q,\nu)$,
is given by

\be F_{A}(q,\nu)=2\pi\int\displaylimits_{E_{min}}^{E_{max}(q,\nu)}
d\!E\int\displaylimits_{k_{min}(q,\nu,E)}^{k_{max}(q,\nu,E)} k\,d\!k\,
P(k,E)
 \label{fnu}
\ee
(assuming $P^p=P^N\equiv P$).
In  analyzing  quasi-elastic scattering in terms of
$y$ scaling \cite{west} a new variable $y=y(q,\nu)$ is introduced and
 Eq. (\ref{fnu})  expressed  in terms of $q$ and  $y$  rather than $q$ and
$\nu$.

The most commonly used  scaling variable~\footnote{
For other types of scaling variables see \cite{rinat}} is obtained \cite{ciofi} starting from
relativistic energy conservation, Eq. (\ref{consen}),
and setting $k=y$,
 $\frac{{\bf k}\cdot {\bf q}}{k q} = 1$, and, most importantly, the
  excitation energy
$E_{A-1}^{\ast}=0$; in other words, $y$ is obtained from the following equation:

\begin{equation}
 \nu + M_{A} = [M_{A-1}^2
 + {y}^2]^{1/2} + [M^{2} + ({y} + {q})^{2}]^{1/2}
 \label{yold}
\end{equation}

 \noindent In this case, $y$  therefore   represents the  longitudinal momentum of
 a
nucleon having the  minimum  removal energy
$(E = E_{min}$, i.e. $E_{A-1}^{\ast} = 0)$. It can be  shown \cite{ciofi} that, at
high values of $q$, $E_{max}  \simeq \infty$ and $k_{min}
\simeq |y-(E-E_{min})|$, so that Eq. (\ref{fnu}) reduces to
\be
F_{A}(q,y) \to f_{A}(y)=2\pi\int\displaylimits_{E_{min}}^{\infty}
d\!E\int\displaylimits_{|y-(E-E_{min})|}^{\infty} k\,d\!k\, \spett
\label{fasymp} \ee
explicitly showing scaling in $y$.
 By defining an experimental scaling function
\be
F_{A}^{exp}(y,q)=\frac{(\sezqe)^{exp}}{(Z\sigma_{ep}+N\sigma_{en})
\frac{E_{p}}{q}}
 \label{fexp}
\ee
 and  comparing it with the theoretical expression, Eq.
(\ref{fnu}), important information can in principle be obtained. For example, from
deviations between Eqs. (\ref{fnu}) and (\ref{fexp}), one can learn about FSI whilst, if scaling
is observed,  one can learn about the nucleon spectral function, ${\spett}$. In practice,
binding effects, i.e. the dependence of $k_{min}$  upon $E$ (and, therefore, $E_{A-1}^{\ast})$,
do not permit a direct relationship between $F_{A}(y)$  and
the longitudinal momentum distributions given by
\be
 f_{A}(y) = 2 \pi
\int\displaylimits_{|y|}^{\infty}
          k\, d\! k \, n_{A}(k)
  \label{fy}
\ee
 where $n_{A}(k)=\int\displaylimits_{E_{min}}^{\infty}d\! E
            P_A(k,E)$ is the nucleon momentum distribution. Even at high momentum transfer,
the contribution of FSI  can "scale" due to the constant value of the
 total $NN$ cross section, thereby confusing a direct extraction of $n_{A}(k)$\cite{ji}. Moreover
it should be pointed out that, when expressed in term of the usual scaling variable, $y$, a
 comparison between experimental and theoretical scaling functions
requires
knowledge  of the nucleon spectral function. This  is   difficult
 to calculate theoretically; on the other hand, theoretical knowledge of nucleon momentum
distributions,
$n_{A}(k)$, is
 rather well known, although  experimentally it is
 very poor.  There are, therefore,   excellent
 reasons to justify an approach to  $y$-scaling based on
 longitudinal
 momentum distributions, $f_{A}(y)$ (Eq. \ref{fy}), rather than the asymptotic scaling
 function,  Eq. (\ref{fasymp}). Apart from the trivial case of the
 deuteron, for which, by definition, $F_{D}(y)=f_{D}(y)$, the problem for complex
  nuclei is that
the  final spectator $(A-1)$ system
 can be left
 in all
possible excited states,  including the continuum. For this reason,
the scaling variable
defined by Eq. (\ref{yold})   can only be identified with  the longitudinal momentum
for   weakly bound, shell model nucleons ( where ${E_{A-1}^\ast}\simeq 0-20 MeV$) but not
for  strongly bound, correlated nucleons (where ${E_{A-1}^\ast} \sim
50-200 MeV$), whose contribution almost entirely saturates
the scaling function at large values of $y$.
Since the definition of the scaling variable is not  unique, it is
prudent to incorporate the most important
     dynamical effects, such as binding corrections, into its
 definition
 in order to establish a global
 link between
experimental data  and longitudinal momentum components. With this in mind we recently introduced
such a scaling variable   which has,  to a large extent,
the desired property of equally well
representing  longitudinal momenta of both weakly and strongly bound nucleons\cite{CW}.
It is based upon the idea  of effectively including in Eq.
(\ref{yold}) the excitation energy of the  $(A-1)$ system due to nucleon-nucleon
correlations. This is given
by \cite{CS,frank}

\begin{equation}
E_{A-1}^{\ast}({\bf k}) = \frac{A-2}{A-1}{\frac{1}{2M}}[{\bf {k}} -\frac
{A-1}{A-2}{\bf {K}}_{CM}]^2
\label{estar}
 \end{equation}

 \noindent  where  $\bf k$ is the relative momentum of a correlated pair
    and ${\bf K}_{CM}$ its CM momentum. This is in good  agreement with  results of  many-body
 calculations  for
nuclei ranging
 from $^3He$ to nuclear matter \cite{manybody}.
We have evaluated   the expectation value of Eq.~(\ref{estar})
using realistic spectral functions obtaining
\footnote{This is slightly different from the form given in ref.\cite{CS} and used in
ref.\cite{CW} where a  term quadratic  in ${\bf k}$ was used instead
of  ${c{_A}} |\bf k|$; both forms are equally well acceptable.}

 \be
<E_{A-1}^{\ast}(k)>
\simeq\frac{A-2}{A-1}{\frac{1}{2M}}{\bf {k}}^2 +{ b_A}-{ c{_A}} |\bf k|
\label{estarav}
 \ee

\noindent The parameters $b_A$ and $c_A$, which
 result from
 the $CM$ motion of the pair, have values ranging  from $17 MeV$ to $43
 MeV$
 and $6.00 \times 10^{-2}$ to $8.00 \times 10^{-2}$, for $^3He$ and nuclear
matter, respectively. The idea was,
therefore, to obtain the scaling
variable by using this in the energy conservation equation,
 Eq. (\ref{consen}), thereby  obtaining:
\begin{equation}
 \nu + M_{A} = [(M_{A-2} + M +
E_{A-1}^{\ast}(y))^{2} + {y}^2]^{1/2} + [M^{2} + ({y} + {q})^{2}]^{1/2}
\label{defycw}
\end{equation}
\noindent In order to
  ensure a smooth transition
between the high and the low values of $y$, we shift the arbitrary scale of $<E_{A-1}^{\ast}(k)>$
in Eq. (\ref{estarav}) by
the average shell-model removal energy,  $<E_{gr}>$: $<E_{A-1}^{\ast}(k)> \rightarrow
<E_{A-1}^{\ast}(k)> - <E_{gr}>$. Note that
$<E_{gr}>$ is not a free parameter since it is obtained from the Koltun sum rule.
Furthermore,
 we use in Eq. (\ref{estarav}) the relativistic form  $\sqrt{M^{2}+{\bf{k}}^2}
 -M$ in place of
${\frac{1}{2M}}{\bf{k}}^2$.




For a heavy nucleus, where  $M_{A-2} + M + E_{A-1}^{\ast}(y) \gg
{y}$, the equation defining the new scaling variable therefore becomes

\be
\sqrt{M^{2}+(y+q)^{2}}+ \sqrt{M^{2}+y^{2}}+{c_A}y = \nu+M-E_{th}-b_A+<E_{gr}>
 \label{defy}
\ee

\noindent where $E_{th}=M_{A-2}+2M-M_{A}$. Disregarding  terms of
order ${c_{A}}^2$ and  $c_{A}/q$,  Eq. (\ref{defy})
 can be solved to obtain the  new scaling variable
 in the form\footnote{These approximations, including the use of  the relativistic form
in  Eq. (\ref{estarav}), have been
 checked
 numerically and found to be
 very good ones in all kinematical regions of interest.}

\be
y_{CW}= -{{\tilde q}\over 2} + \left [{{\tilde q}^2\over{4}} - \frac{4
{\nu_{A}}^2 M^{2}-\rm{ W_A}^{4}}{\rm{4\, W_A}^{2}}\right ]^{1/2}
\label{ycw3}
 \ee

\noindent where $\nu_A=\nu+2M-E_{th}-b_A+<E_{gr}>$, $\tilde{q}=q+c_A\nu_A$ and
 $\rm{W_A}^{2} \equiv
{\nu}_A^2 - {q}^2$.

It is worth emphasizing that, at low values of $y$,
the usual scaling variable is recovered, $y_{CW}\simeq y$; indeed, $y$ can
 be obtained from Eq.
(\ref{ycw3}) as the limiting case where
$b_A=c_A=<E_{gr}>=0$.  Furthermore, for the deuteron ($A = D$),
$b_D=c_D=E_{gr}=0$, $E_{th}=|\epsilon_D|=2.225 MeV$, $\nu_D=\nu+M_D$,
${\tilde q}=q$,  and $\rm{W_A}^{2} \equiv {\nu_{D}^2 - q^2}=
{(-Q^2+M_{D}^2+2M_{D}\nu)}$, leading to  the usual  deuteron scaling variable
$y_{CW}= y_{D}= -{{q} \over 2} + \left [{{ q}^2  \over{4}} - \frac{4
{\nu_{D}^2} M^{2}-\rm{ W_D}^{4}}{\rm{4{W_D}}^{2}}\right ]^{1/2}$. The use of
 $y_{CW}$ instead of the
usual $y$
has the following advantages:
\begin{enumerate}

\item Since, at large values of $q$, the limits on the integrations in Eq.
(\ref{fnu}),  $E_{max}
\simeq E_{min}
 \simeq \infty$
and  $k_{min} |\simeq y_{CW}|$ (instead of  $k_{min}= |y-(E-E_{min})|$),
the  asymptotic scaling function when expressed as a function of $y_{CW}$,
  directly measures the
longitudinal momentum distributions:
\be
F_{A}(y_{CW}) \simeq f_{A}(y_{CW}) = 2 \pi
\int\displaylimits_{|y_{CW}|}^{\infty}
          k\, d\! k \, n_{A}(k)
\label{fynew}
\ee
 \noindent Thus, plotting the data in terms of $y_{CW}$ can
 provide direct access to the nucleon momentum distributions.
\item Since many body calculations \cite{manybody} show that at high momenta,
 $k \geq
1.5-2 fm^{-1}$, all nuclear momentum distributions are simply  rescaled versions of the deuteron,
\be
 n_A(k)\cong C_A n_D(k)
 \label{nak}
 \ee
   where $C_A$ is a  constant, one should also
   expect
   \be
 F_A(q,y_{CW})\cong C_A F_D(q,y_{CW})
 \label{fak}
 \ee
\noindent On the other hand  no such proportionality  is
expected between $F_A(q,y)$ and
 $F_D(q,y)$.
\item By eliminating  binding effects, scaling violations observed
in the experimental data,
 Eq. (\ref{fexp}),
can thereby be ascribed  to the FSI, allowing a relatively clean separation between the two
scale violating effects.

\end{enumerate}

In order to check the validity of the above points,
we show in Figs. 2 and 3 the experimental scaling functions for $A=2, 4$ and $56$,
 plotted in terms of the old and new scaling variables, respectively.
 It can be seen from Fig. 2 that  the scaling functions, $F_A^{exp}(y,q)$ for $A=4$ and $56$,  do
not exhibit any simple  proportionality to the deuteron scaling function  at large values of
$|y|$, in contrast to the case of
$F_A^{exp}(y_{CW},q)$,
 shown in Fig. 3, which agrees remarkably well with the predictions of
    Eq. (\ref{fak}). The $Q^2$-dependence of $F_A^{exp}(y_{CW},q)$,
 is shown in Figs. 4 and 5;  in the latter figure we show  the same scaling function divided by the
constant $C_A$ of Eq. (\ref{nak}), taken
 refs. \cite{CS,manybody}.

\section{The final state interaction}

Figs. 4 and 5 show an approach to scaling from above and represent a clear signature
of the effects of FSI. These were taken into account by a method \cite {faralli}
similar
to the one used in Ref. \cite{CS}: both the rescattering of the struck nucleon
via an optical potential generated by the shell model $(A-1)$ spectator  system,
as well as the two-nucleon rescattering in the final state when the struck nucleon is a partner
of a correlated pair, were taken into account. The results are exhibited in
Fig. 6 by the continuous line, while the dashed line represents the PWIA results and
the dotted line the longitudinal momentum distributions. The following remarks are
in order:
\begin{enumerate}
\item At high $Q^2$ the PWIA result is very similar
to $f_A(y_{CW})$,  as expected from Eq. (\ref{fynew}). This is in marked contrast to what happens
in the usual approach to $y$-scaling (cf. Fig. 2);
\item The calculated FSI {\it decreases} with $Q^2$,  {\it approaching} the PWIA
 limit from above and, more
importantly, agrees fairly well  with the trend of the
data.
\end{enumerate}

The latter point requires specific comments, for it is a common belief that at
high $Q^2$ FSI should be governed by the total
nucleon-nucleon (NN) cross section which exhibits a constant behaviour for
$p_N \geq 1.2 GeV/c$, where $p_N$ is the lab momentum of the incident
nucleon (see ref. \cite{baldini}). In treating this point a crucial role is played by the
 four-momentum, $p^{\prime}$, of the struck nucleon after the absorption
  of the virtual photon. The usual procedure  is to
  approximate  its kinetic
energy  before rescattering by

\be
T_q=E_q-M=\sqrt{q^2+M^2}-M \simeq \frac{Q^2}{2M}
\label{tq}
\ee
\noindent Such an approximation should be  reasonable for $y \simeq 0$ $(x
\simeq 1)$,  where the struck nucleon in the final state
 can be regarded as  quasi-free (almost on
shell, $p_{1}^2\simeq M^2$). It should, however, be questioned at high negative values of $y$
($x\gg 1$), where, after absorbing the photon,  the struck nucleon is  far
 off-shell with invariant mass
 \be
  p_{1}^2\simeq {M^2} + Q^2({1 \over x}-1)- {\bf k}^2- 2k_{L}|{\bf q}|
  \label{invariantm}
  \ee
  where $k_{L}\equiv {\hat q}|{\bf k}|$.  As a result,
  one has to consider the nucleon-nucleon cross section for a far-off-shell
incident nucleon. This  is a very difficult task. However,
at the very least, one should
consider off-shell kinematics.
In such a case,  the CM kinetic energy, $T_{off}$, of  a two-nucleon
 pair after one nucleon
 has been
struck by the virtual photon
but  before it  rescatters  from a
spectator nucleon is,
 at large negative  $y$ ($x \gg 1$),  {\it less} than the inelastic threshold
  (= $m_{\pi}$)
.
 Thus, it is mostly the  {\it elastic}
NN cross section, which {\it decreases} with energy, that must be used rather than the constant
total NN cross section.

\section{Extraction of the nucleon momentum distributions}
Given the approach of the scaling function $F_A(y_{CW},q)$
 to the PWIA result exhibited in Figs. 4-7, the longitudinal
momentum distribution can be extracted from the experimental data without the
 uncertainties associated with the subtraction of the so called binding
 correction (see Refs. \cite{ciofi}, \cite{rinat}).
 The example of $^{56}Fe$ is given in Fig. 8. By taking the derivative
 of $f(y_{CW})$ the nucleon momentum distributions $n(k)$ can be
 obtained.
\section{Acknowledgments}
 CCdA and GBW acknowledge the hospitality of the Santa Fe Institute where
 part of this work was   completed.




\begin{figure}[ptb]
\vspace*{2cm}
\centerline{\psfig{file=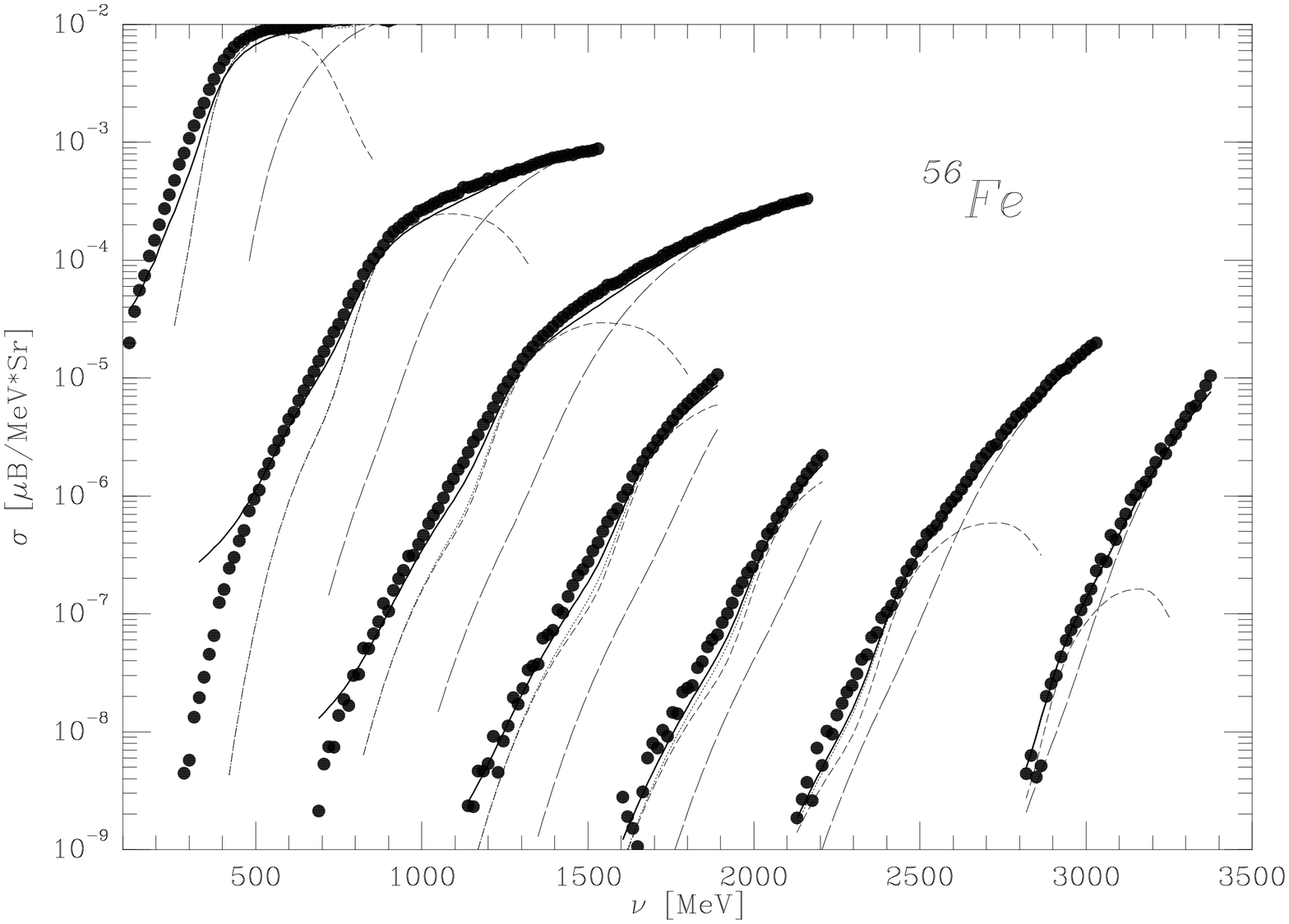,height=10cm,angle=270}}
\vspace{2cm}
\caption{The experimental data from Ref. 1 for $\fec$ (solid circles)
compared to theoretical calculations. {\it Short-dashes}: quasi-elastic
PWIA; {\it Long-dashes}: inelastic PWIA; {\it Dots}: sum of quasi-elastic and
inelastic PWIA. The continuous line includes the effects of FSI. The
various sets of experimental data correspond  to different values of
the scattering angle $\theta$, ranging from $15^o$ to $74^o$, from left
to right
.}
\label{fig1}
\end{figure}
\clearpage

\begin{figure}[tb]
\vspace*{2cm}
\centerline{\psfig{file=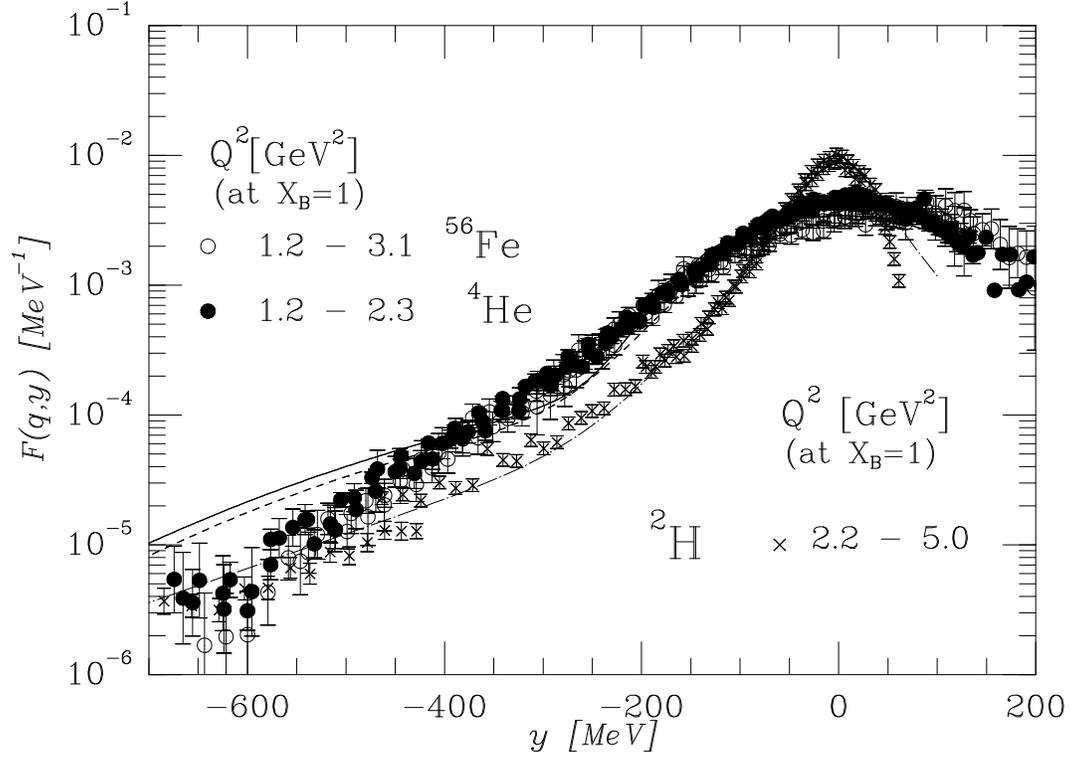,height=10cm,angle=270}}
\vspace{2cm}
\caption{The experimental scaling function of $\heq$ (solid circles)
and $\fec$ (open
circles) plotted
{\it vs}
 the usual  scaling variable
$y$, compared to the scaling function of  $^2H$ (crosses).
The  solid,
 dashed and dot-dashed curves
 represent the theoretical  longitudinal momentum distributions of $\fec$,
  $^4He$ and
$^2H$,
respectively
(after Ref.2)
.}
\label{fig2}
\end{figure}
\clearpage

\begin{figure}[tb]
\vspace*{2cm}
\centerline{\psfig{file=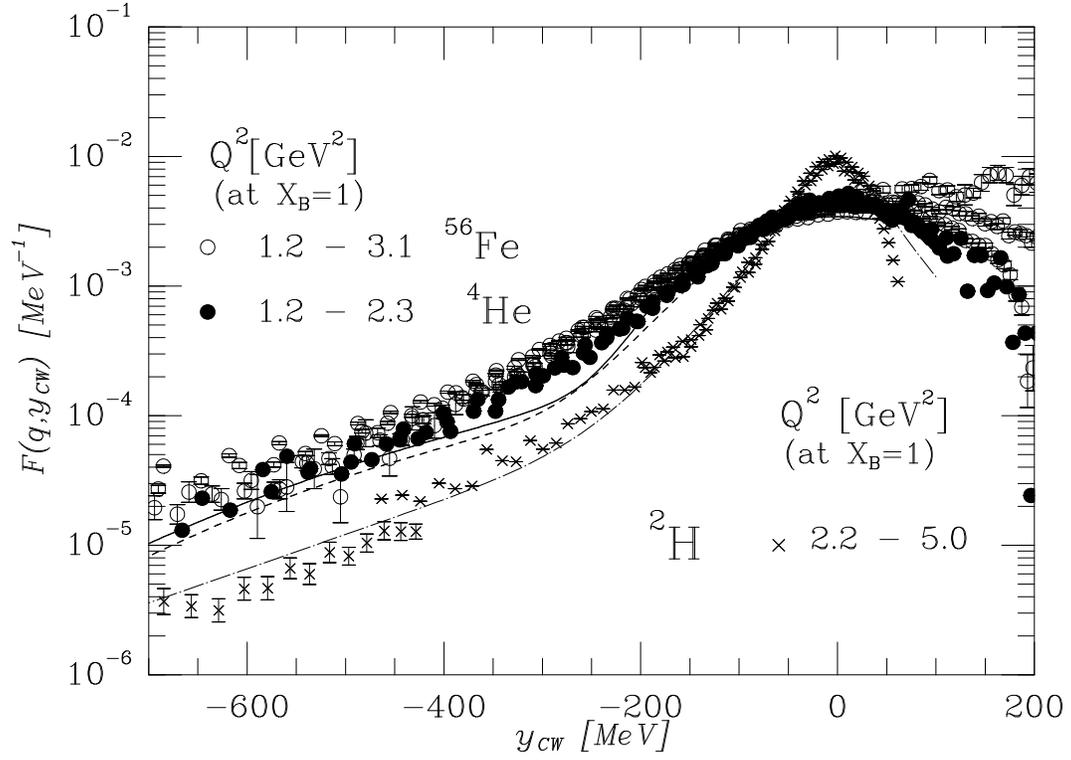,height=10cm,angle=270}}
\vspace{2cm}
\caption{The same as in Fig. 2 but plotted {\it vs.} $y_{CW}$ as
defined by Eq.(\ref{defy}), (after Ref.2 )}
\label{fig3}
\end{figure}
\clearpage

\begin{figure}[tb]
\centerline{\psfig{file=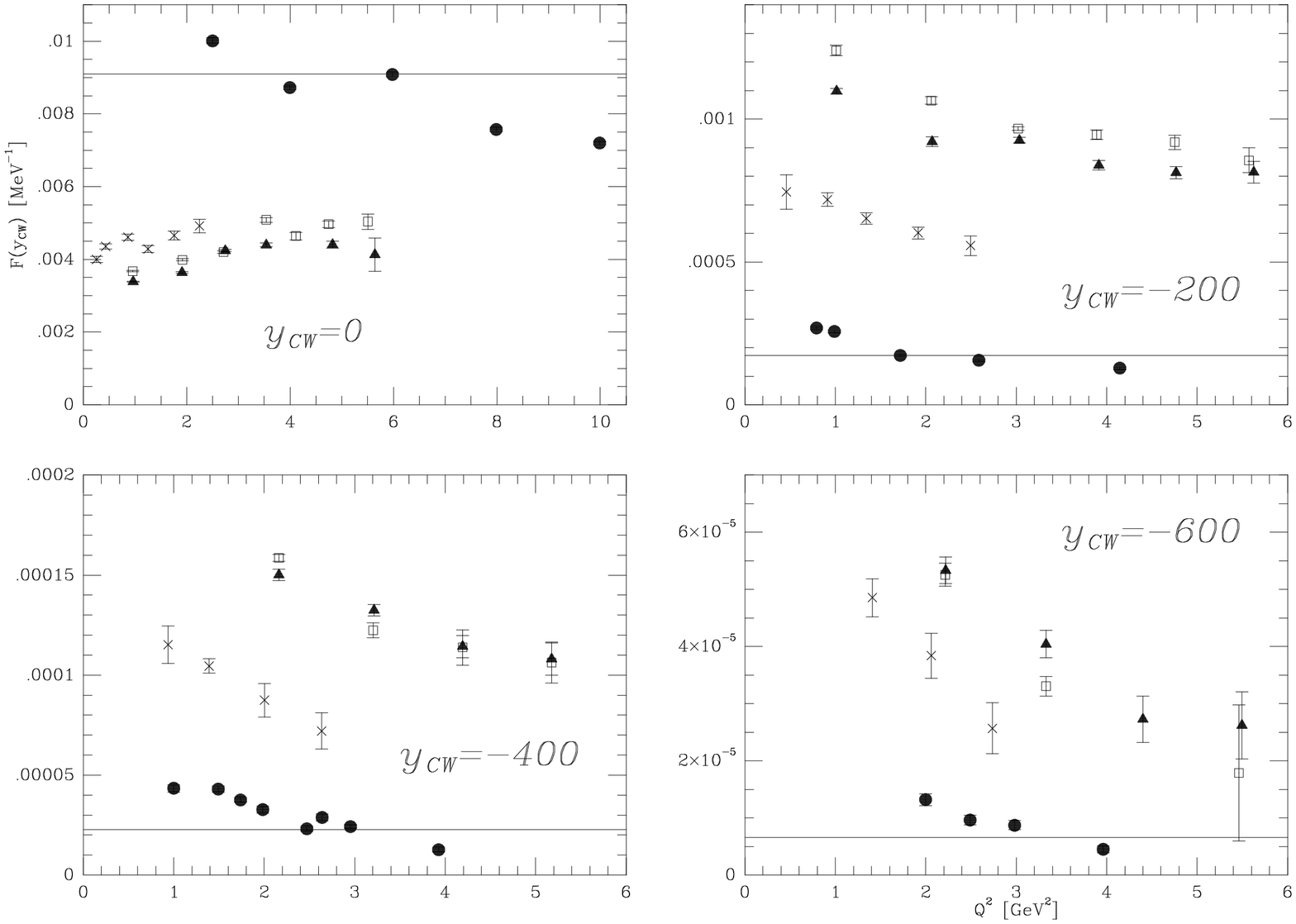,height=12cm,angle=270}}
\vspace{2cm}

\caption{The scaling functions  of $\hd$ (solid circles), $\heq$ (crosses), $\cd$
 (triangles)
 and $\fec$
(squares) at fixed
 values of $y_{CW}$ plotted {\it vs} $\qud$.}
\label{fig4}
\end{figure}
\clearpage

\begin{figure}[tb]
\centerline{\psfig{file=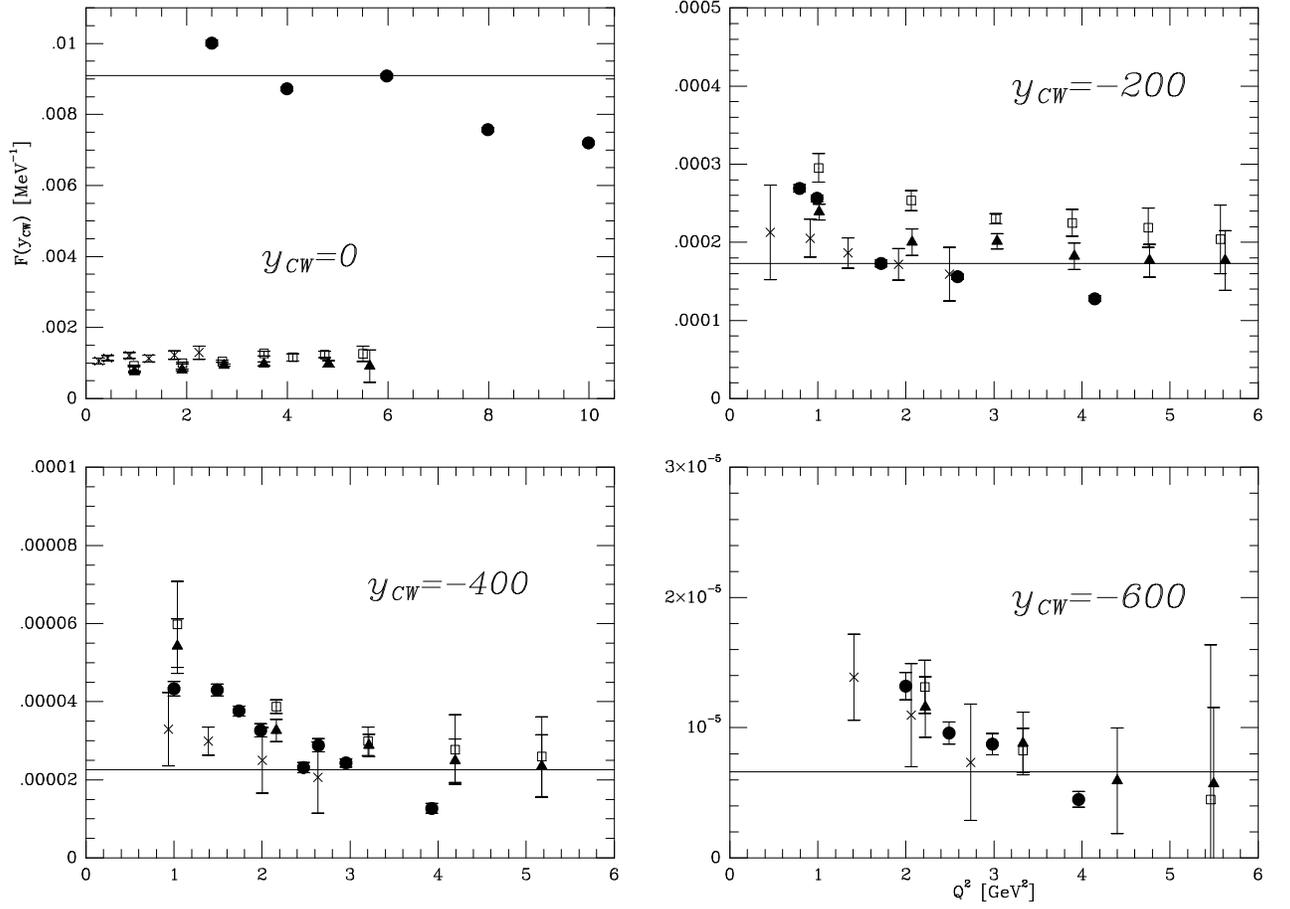,height=12cm,angle=270}}
\vspace{2cm}
\caption{The same as in Fig. 4 but with $F(y_{CW},q)$ divided by the constant, $C_A$, defined in
Eq. (\ref{nak}). The data exhibit a universal
behaviour where  the scaling  function of any nucleus in a wide range of  $y_{CW}$
is simply $C_A$ times that of the deuteron.}
\label{fig5}
\end{figure}
\clearpage

\begin{figure}
\vspace*{2cm}
\centerline{\epsfysize=15cm \epsfbox{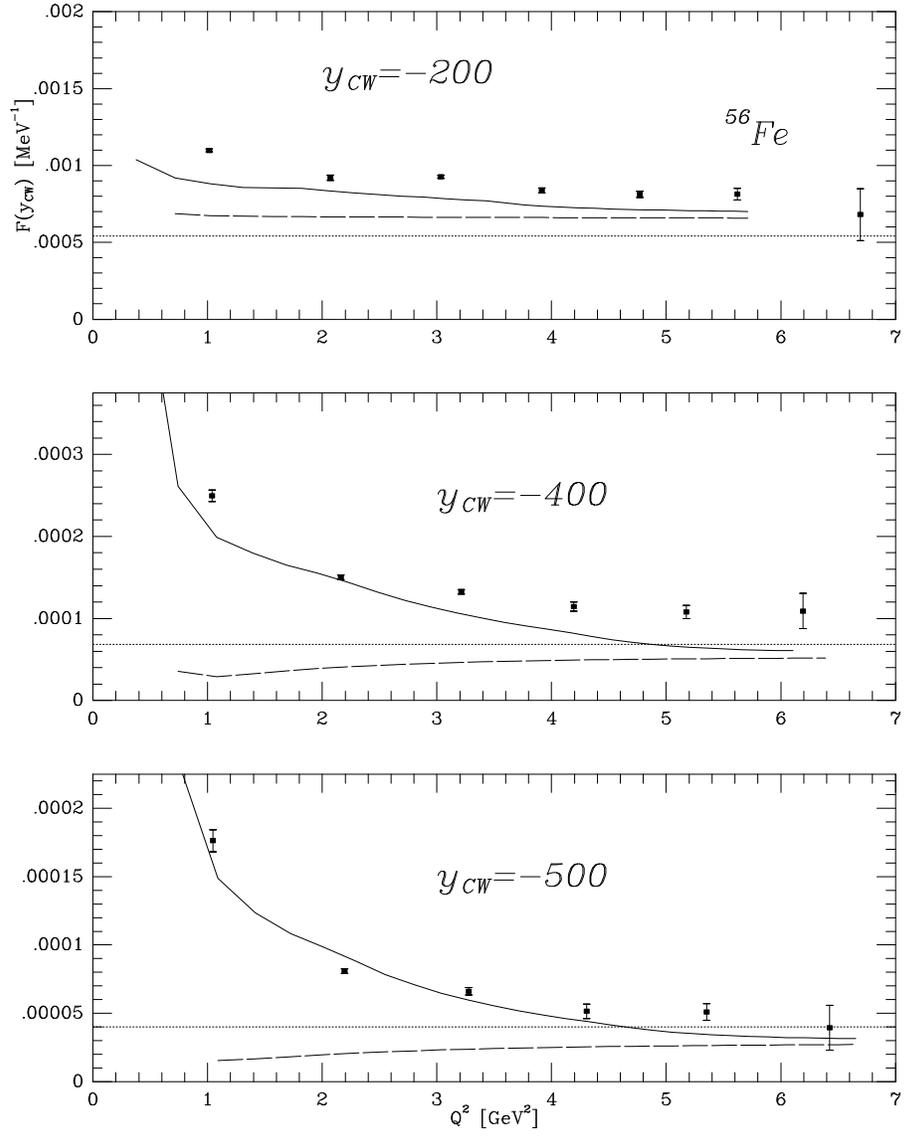} }
\vspace{2cm}
\caption{The scaling function of  $\fec$ {\it vs} $\qud$ for fixed values of
$y_{CW}$  compared with the PWIA (dashed line) and the full FSI result
(full line). The dotted line represents the longitudinal momentum distributions.}
 \label{fig6}
\end{figure}
\clearpage


\begin{figure}
\vspace*{2cm}
\centerline{\epsfysize=15cm \epsfbox{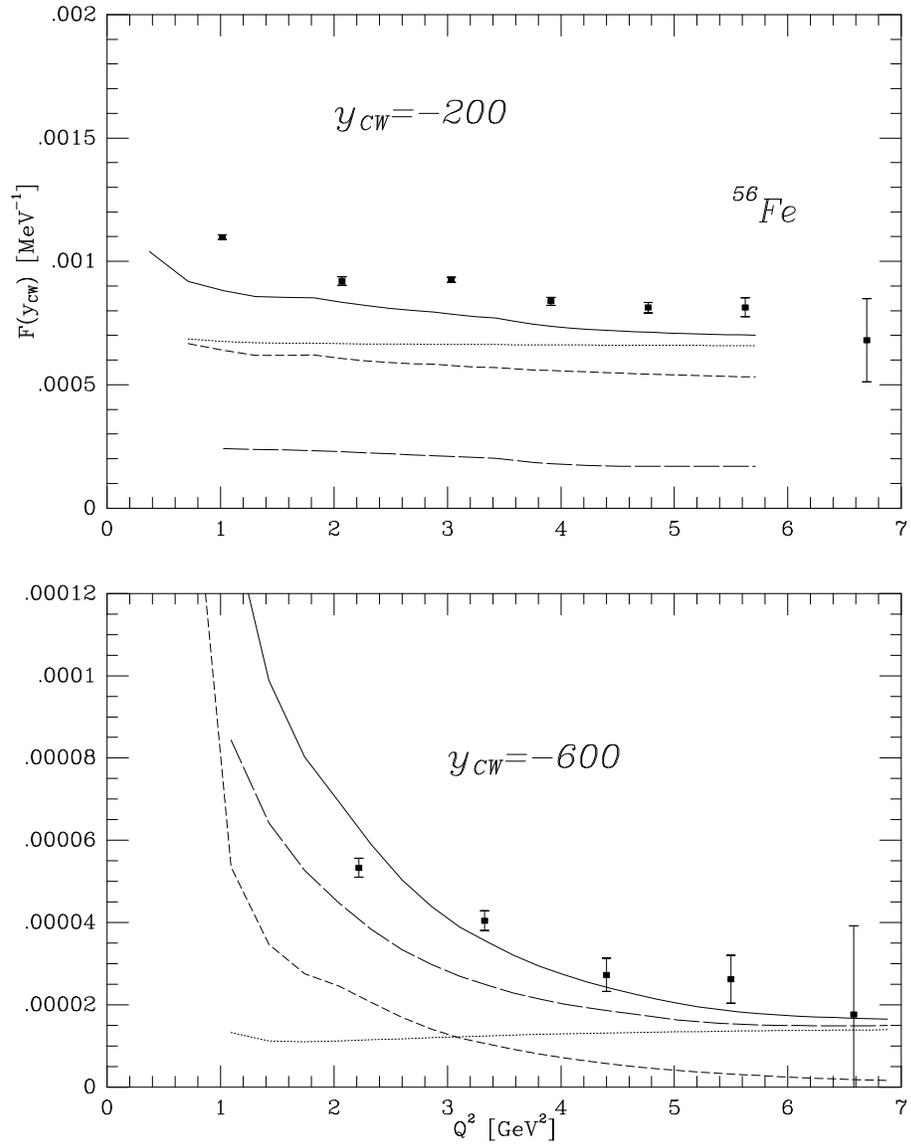} }
\vspace{2cm}
\caption{The various contributions to the full FSI result (full
curve): {\it PWIA}: dots;
{\it optical potential}: short dashes; {\it two-nucleon rescattering}: long dashes.}
 \label{fig7}
\end{figure}
\clearpage

\begin{figure}[tb]
\vspace*{2cm}
\centerline{\psfig{file=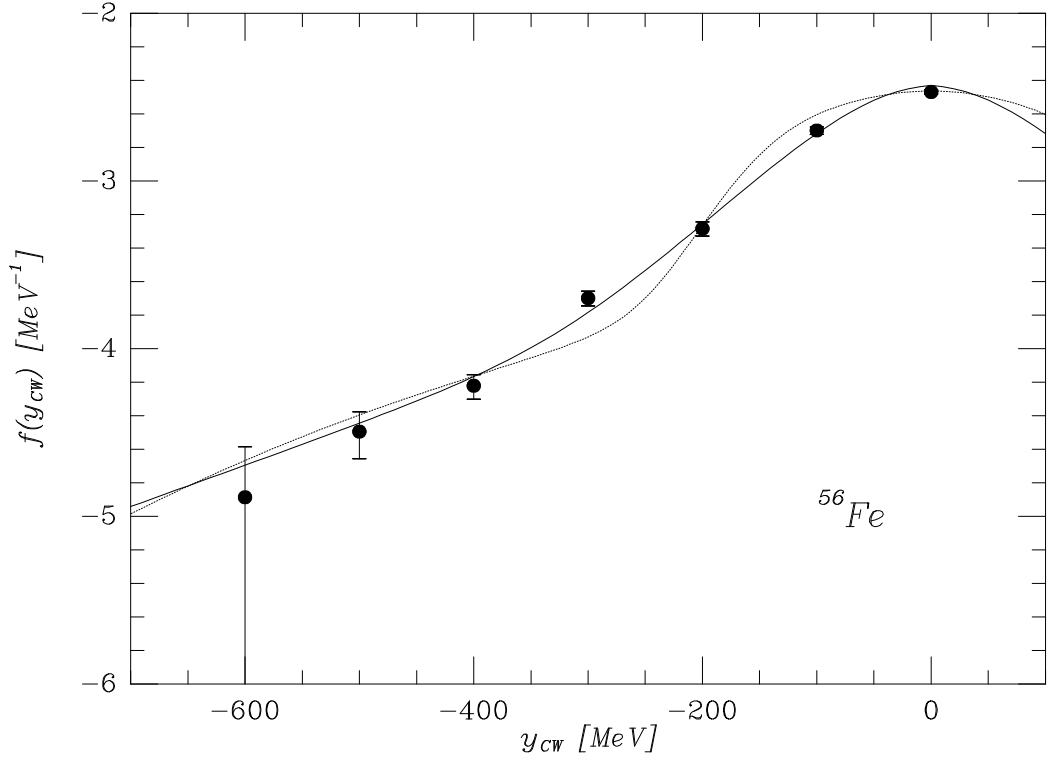,height=10cm,angle=270}}
\vspace{2cm}
\caption{The longitudinal momentum distribution (dots) for
$\fec$ obtained from the results shown in Figs. 4-7. The dotted and solid
curves correspond to two different theoretical longitudinal momentum
 distributions.}
  \label{fig8}
\end{figure}
\clearpage

\end{document}